\documentclass{iopconfser}
\usepackage{subcaption}

\usepackage{natbib}
\usepackage{enumitem}
\usepackage{graphics}
\usepackage{tikz}

\begin{document}
\title{Search for Neutrino Counterparts to Gravitational Wave Events in Super-Kamiokande during the LIGO/Virgo/KAGRA O4 Observing Run}

\author{L. N. Machado$^{1}$, on behalf of the Super-Kamiokande Collaboration}

\affil{${}^1$School of Physics \& Astronomy, University of Glasgow, Glasgow, United Kingdom}

\email{lucas.nascimentomachado@glasgow.ac.uk}

% \newpage
\begin{abstract}
In recent years, the Super-Kamiokande (SK) experiment has contributed to multi-messenger astronomy by searching for neutrino signals in coincidence with gravitational wave events.
The fourth observing run (O4) of the LIGO/Virgo/KAGRA collaborations started in May 2023. During the initial phase, known as O4a, LIGO identified 81 gravitational wave candidates.
Neutrino follow-up searches with O4a were performed in SK. While no significant signal was observed, flux limits were computed.
\end{abstract}

\section{Neutrinos in Multi-Messenger Era}
Multi-messenger astronomy combines information from different cosmic messengers to reveal the Universe in ways a single messenger cannot. Neutrinos play a key role in this era: they originate from astrophysical sources such as supernovae; they interact only weakly, escaping dense regions opaque to light; and they can point back to their sources. Neutrinos have already had a major impact with the discovery of neutrino oscillations~\citep{ALIANI2002393}, and with the first neutrino detection from a core-collapse supernova from SN1987A~\citep{TOTSUKA1988189}.

Experiments such as IceCube and KM3NeT/ARCA are pioneers in the high-energy search for neutrino signals. IceCube has measured the first diffuse astrophysical flux up to PeV energies~\citep{icecube}, and KM3NeT/ARCA reported record-breaking $\sim$220 PeV neutrinos in 2025~\citep{km3net}. The next generation of detectors such as IceCube-Gen2, Hyper-Kamiokande, DUNE, and RNO-G, will strengthen the synergy between neutrinos and other cosmic messengers.

\subsection{Super-Kamiokande (Super-K)}
Super-K is a 50‑kton water Cherenkov detector in Japan, operational since 1996~\citep{Fukuda_2003}. In 2020, Super-K started the SK-Gd phase with the loading of $\mathrm{Gd_2(SO_4)_3 \cdot 8H_2O}$ to enhance the sensitivity to low energy $\bar{\nu}_e$ by improving the detection efficiency of inverse beta decay interactions~\citep{Beacom_2004}. Super-K plays a key role in multi-messenger astronomy:
\begin{itemize}\setlength\itemsep{0.01em}
    \item Sensitive to MeV–GeV neutrinos from diverse astrophysical sources.
    \item Operates an active supernova alarm system~\citep{Super-Kamiokande:2024pmv}, issuing supernova alerts via GCN/SNEWS.
    \item Provides an open pre-supernova alert system (in collaboration with KamLAND)~\citep{Super-Kamiokande:2022bwp, KamLAND:2024uia}.
    \item Performs follow-up searches triggered by gravitational wave (GW) events.
\end{itemize}

Super-K is implementing a real-time system for a faster response to search for neutrino counterparts to GW events, which could help improving the source sky localization and sensitivity.

% \subsection{Astrophysical Neutrino Sources}
% Compact binary mergers are prime candidates for neutrino–GW coincidence searches. 
% \begin{itemize}
%     \item Neutron Star–Black Hole (NSBH) mergers: may produce both high-energy and thermal neutrinos, depending on remnant and ejecta.
%     \item Binary Neutron Star (BNS) mergers: emit low-energy neutrinos (MeV) within milliseconds of the GW signal.
%     \item Black Hole–Black Hole (BBH) mergers: no neutrino emission expected, but some models allow for high-energy neutrino production if matter is present (e.g.accretion disk).
% \end{itemize}

% Low-Energy Neutrinos, in the order of MeV, are expected from BNS and NSBH mergers, emitting thermal MeV neutrinos within < 1~s of merger.
% Simulations predict all neutrino flavors, with $\bar{\nu}_e$ typically dominating. The peak luminosity $> 10^{53}$ erg/s, with neutrino energies ranging from 5 to 30 MeV. 

% High‑Energy Neutrinos, in the order of GeV to PeV, are produced by  non-thermal neutrinos processes, through jet and shock acceleration:
% Internal, collimation, and reverse shocks can yield GeV–PeV neutrinos. Mildly relativistic jets (e.g. with neutron–proton conversion) favor production in the GeV–TeV range. Extended PeV–EeV emission possible from ejecta–wind interactions or long-lived remnants.

\subsection{Astrophysical Neutrino Sources}
Compact binary mergers are the main candidates for neutrino–GW coincidence searches:

\begin{itemize}\setlength\itemsep{0.01em}
    \item Neutron Star–Black Hole (NSBH) mergers may produce both high-energy and thermal neutrinos (low-energy), depending on remnant state and ejecta.
    \item Binary Neutron Star (BNS) mergers generate thermal neutrinos, dominated by $\bar\nu_e$, within milliseconds of the GW signal. Peak luminosities exceed  $10^{53}\,\mathrm{erg/s}$, with energies ranging from 5~to~30~MeV~\citep{refId0}.
    \item Generally for Black Hole–Black Hole (BBH) mergers, neutrino emission is not expected, unless matter is present (e.g., in AGN disks), where jet–disk interactions can produce GeV–PeV neutrinos~\citep{Zhou_2023}.
\end{itemize}

\section{O4 follow-up searches in Super-K}
Several Super-K searches for neutrino counterparts to GW events have been conducted~\citep{Abe_2016,Abe_2018,Abe_2021}, with no significant coincidence observed so far. The fourth observing run of the LIGO/Virgo/KAGRA collaborations started in May 2023, with the first period (O4a) ending in January 2025. For this follow up search, only the first 56 significant GW events were considered: two likely NSBH mergers, and the majority BBH mergers.

Events in Super-K are classified as low-energy (LE, 2.5–100 MeV), fully contained (FC, 0.1–10 GeV, $\Delta\theta \sim 10^{\circ}$–$100^{\circ}$), partially contained (PC, 0.1–100 GeV, $\Delta\theta \sim 10^{\circ}$), and upgoing muons (UPMU, 1.6–1000 GeV, $\Delta\theta \sim 3^{\circ}$–$10^{\circ}$). LE events follow the selection used in diffuse supernova neutrino background (DSNB) and solar analyses.

For the search in the high energy samples (FC, PC, UPMU), a flat background was assumed over the full period. Two test statistics were performed: a time correlation ($p_{time}$) and directional test ($p_{lambda}$). For the search in the low energy samples (DSNB, solar), the background was estimated from a four-day window around the GW time, excluding search window. Only time correlation test was performed. No significant excess was observed in both samples. Tables~\ref{tab:atmpd_results}, \ref{solar_tab}, ~\ref{dsnb_tab} show the observed events in coincidence with GW-O4a, only in cases where at least one sample contained one or more events.
\begin{table}[htbp]
\centering
\footnotesize
\begin{tabular}
{|c|c|c|c|c|c|c|c|c|c|}
\hline
GW  EVENT &	GW  TYPE	& FC EXP  &	FC OBS	&PC EXP&	PC  OBS	&UPMU EXP&	UPMU  OBS&	$p_{time}$ 	&$p$\\
\hline
S230624av &	BBH (95.3\%)&	1.10E-01&	1&	7.09E-03&	0&	1.39E-02&	0&	0.12&	0.41\\
S230630am&	BBH (98.3\%)&	1.08E-01&	1&	7.08E-03&	0&	1.39E-02&	0&	0.12&	0.53\\
S230731an&	BBH (81.4\%)&	1.08E-01&	1&	7.09E-02&	0&	1.39E-02&	0&	0.12&	0.50\\
S230807f&	BBH (95.3\%)&	1.08E-01&	1	&7.09E-03&	0&	1.39E-02&	0&	0.12&	0.68\\
S230819ax&	BBH (99.3\%)&	1.08E-01	&1&	7.09E-03&	0	&1.39E-02&	0	&0.12&	0.20\\
S230927be&	BBH (100.0\%)&	1.08E-01&	1	&7.10E-02&	0&	1.40E-02&	0	&0.12&	0.54\\
S231001aq&	BBH (99.6\%)&	1.08E-01&	1&	7.10E-03&	0	&1.40E-01	&0&	0.12&	0.99\\
\hline
\end{tabular}
\caption{Events in FC, PC and UPMU samples in coincidence with GW-O4a.}
\label{tab:atmpd_results}
\end{table}

\begin{table}[htbp]
\centering
\footnotesize
% \begin{minipage}[t]{0.4\linewidth}
\begin{subtable}[t]{0.48\textwidth}
\centering
\resizebox{\linewidth}{!}{
\begin{tabular}
{|c|c|c|c|c|}
\hline
GW event&	GW TYPE&	BKG EXP&		N OBS	&$p_{time}$\\
\hline
S230814ah	&BBH (100.0\%)&		9.34E-01&	3&	0.07\\
S230731an	&BBH (81.4\%)&		9.73E-01&	3&	0.08\\
S230904n	&BBH (91.1\%)	&1.02E+00&	3&	0.08\\
S230924an	&BBH (100.0\%)&		8.71E-01&	2&	0.22\\
S230807f&	BBH (86.4\%)&		9.50E-01&	2&	0.25\\
S230726a&	BBH (100.0\%)&		9.80E-01&	2&	0.26\\
S230702an&	BBH (100.0\%)&		1.01E+00&	2&	0.27\\
S230723ac&	BBH (86.7\%)&		1.01E+00&	2&	0.27\\
S230628ax&	BBH (100.0\%) &1.01E+00&	2&	0.27\\
S230601bf&	BBH (100.0\%)&		1.07E+00&	2&	0.29\\
S230529ay&	NSBH (62.4\%)&		1.09E+00&	2&	0.30\\

\hline
\end{tabular}
}
\caption{Solar sample ($6 MeV < E < 8 MeV$).}
\label{solar_tab}
% \end{minipage}
\end{subtable}
% \end{table}
\hfill
% \begin{minipage}[t]{0.4\textwidth}
\begin{subtable}[t]{0.48\textwidth}

% \begin{table}[htbp]
\centering
\footnotesize
\resizebox{\linewidth}{!}{
\begin{tabular}
{|c|c|c|c|c|c|}
\hline
GW event&	GW TYPE &	BKG  EXP&		N OBS	&$P_{time}$ \\
\hline
S230924an&	BBH (98.6\%)&		1.27E-01&	2&	0.007\\
S230729z&	BBH (99.7\%)&		1.25E-01&	1&	0.118\\
S230822bm&	BBH (98.1\%)&		1.47E-01&	1&	0.137\\
\hline
\end{tabular}
}
\caption{DSNB sample ($E > 8 MeV$). Although S230924an shows a low $p$-value, it is not significant once the false-positive rate is taken into account.}
\label{dsnb_tab}
% \end{minipage}
\end{subtable}
\caption{Events in low energy samples in coincidence with GW-O4a.}
\end{table}

% \begin{figure}[htbp]
% \centering
% \begin{subfigure}%{0.48\linewidth}
% \centering
% \includegraphics[scale=0.4]{lowE.png}
% \caption{Low-energy sample.}
% \label{fig:upperlimit1}
% \end{subfigure}
% \hfill
% \begin{subfigure}%{0.48\linewidth}
% \centering
% \includegraphics[scale=0.4]{}
% \caption{High-energy sample.} % required!
% \label{fig:upperlimit2}
% \end{subfigure}
% \caption{}
% \label{fig:upperlimit}
% \end{figure}

\begin{figure}[htbp]
\centering
\begin{minipage}{0.53\linewidth}
\resizebox{\linewidth}{!}{\includegraphics{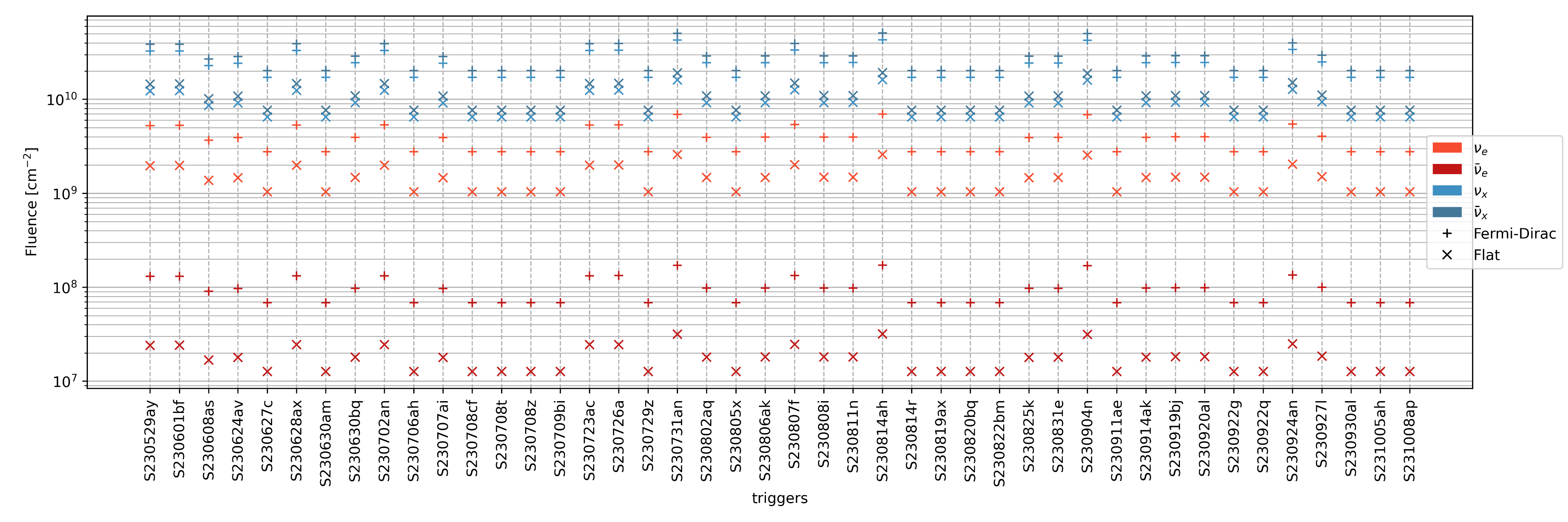}}
\end{minipage}\hfill
\begin{minipage}{0.47\linewidth}
\resizebox{\linewidth}{!}{\includegraphics{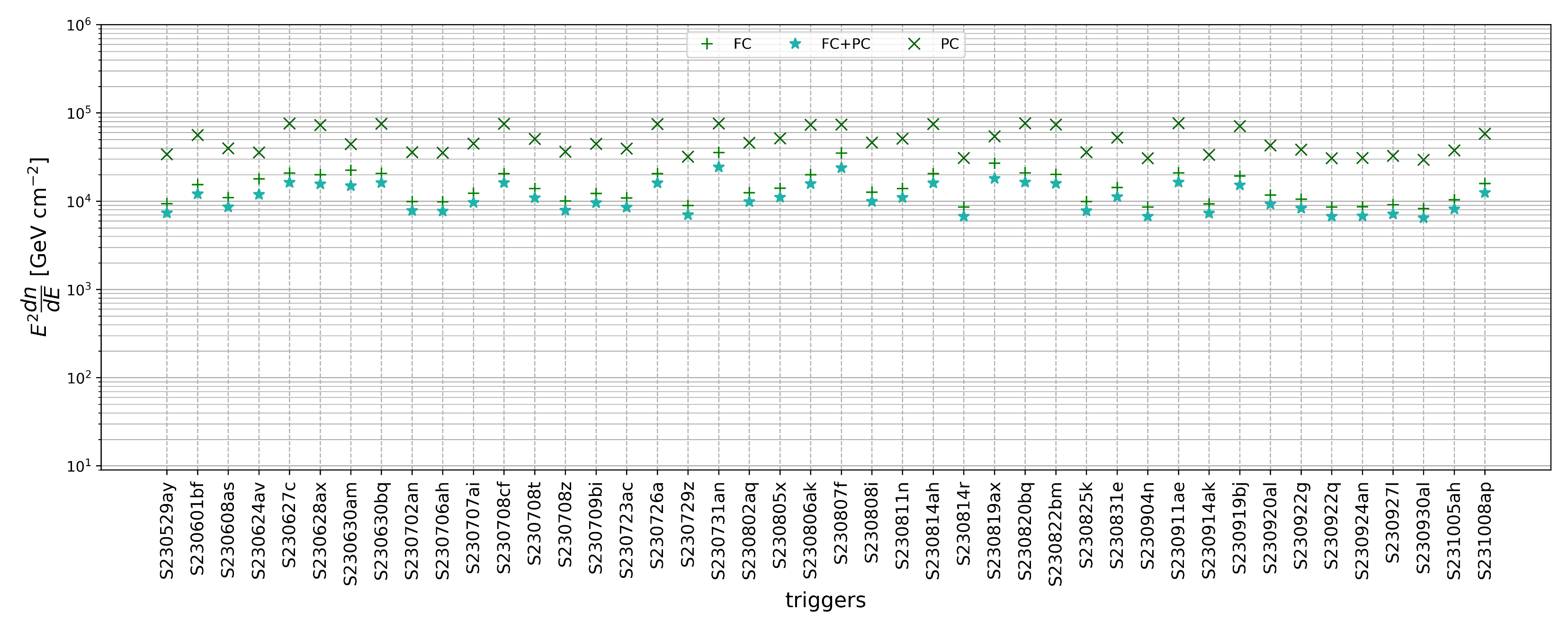}}
\end{minipage}
\caption{90\% C.L. upper limits for all neutrino flavors in coincidence with GW triggers. 
Left: fluence limits from the low energy sample, using both Fermi--Dirac and flat spectra. 
Right: $E^{2} dN/dE$ limits assuming an $E^{-2}$ spectrum, based on FC, PC, and FC+PC samples.}
\label{fig:upperlimit}
\end{figure}

Figure~\ref{fig:upperlimit} displays the derived 90\% C.L. upper limits for neutrinos associated with GW triggers, assuming different spectral models.

\section*{Conclusion and Prospects}
A search for neutrino counterparts for events in the GW-O4a in Super-Kamiokande data was performed, but no significant signal was observed. A full catalog search of neutrino counterparts for O4 will be conducted soon, and the development of a real-time system for follow-up searches is on-going.

% \newpage
% \bibliography{main.bib}

% \newpage
% \appendix
% \section*{Appendix A: Author list}

% \input{authors-20250530}

\end{document}